\documentclass[a4paper]{iopart}

\pdfoutput=1

\usepackage{iopams}
\usepackage{type1cm}
\usepackage[english]{babel}
\usepackage{graphics,graphicx,rotating}
\usepackage[square,sort&compress,numbers]{natbib}
\usepackage[colorlinks]{hyperref}
\usepackage{mathptmx}
\usepackage{enumerate}
\usepackage{bm}

\newcommand{\out}[1]{}

\begin{document}

\title[Analytical meets numerical relativity]{Analytical meets
numerical relativity -- status of complete
gravitational waveform models for binary black holes}
\author{Frank Ohme}
\address{Max-Planck-Institut f\"ur Gravitationsphysik,
Albert-Einstein-Institut, Am M\"uhlenberg 1, 14476 Golm, Germany}
\ead{frank.ohme@aei.mpg.de}

\date{\today}

\begin{abstract}
 Models of gravitational waveforms from coalescing black-hole
binaries play a crucial role in the efforts to detect and interpret
the signatures of those binaries in the data of large-scale
interferometers. Here we
summarize recent models that combine information both from
analytical approximations and numerical relativity. We briefly lay
out and compare the strategies employed to build such complete models
and we recapitulate the errors associated with various aspects of the
modelling process.
\end{abstract}


  \section{Introduction}

The world-wide effort to directly detect
\emph{gravitational waves} (GWs) for the first time is an ambitious
project that unites the expertise from various fields in
experimental and theoretical physics. A network of instruments,
containing the Laser Interferometer
Gravitational-wave Observatory (LIGO)
\cite{Abbott:2007kv,Sigg:2008zz,Smith:2009bx}, VIRGO
\cite{Acernese:2008zzf,Accadia:2011zz} and GEO600
\cite{Grote:2008zz,Luck:2010rs}, will soon reach a sensitivity where
the signatures of coalescing compact binaries are expected to be seen
above the noise level of the detectors a few times to hundreds of
times per year~\cite{Abadie:2010cf}. In the case of binaries
that consist of black holes (BHs) and/or neutron stars, the correct
interpretation of the GW signals crucially depends on the
quality of theoretically predicted \emph{template waveforms} that have
to be used to identify the physical properties of the source.

This paper focuses on waveform families of
\emph{binary BHs} as they constitute one of the most
promising sources of a first direct detection of GWs.  Their
modelling typically combines two very different approximation
procedures. One describes the early inspiral of both objects through
an asymptotic expansion in terms of the relative
velocity $v/c$, where $c$ is the speed of light. As long as this
quotient is small, the resulting post-Newtonian (PN) equations are an
adequate representation of the dynamical evolution of the binary
\cite{lrr-2006-4}.
Because of the simple form of PN approximants that provide the GW
signal in
terms of differential equations or, in some cases, even in a closed
form, they have long been the favourite
tool for data-analysis applications.

However, as the two BHs orbit around each other, they lose
energy through the emission of GWs, and their distance shrinks along
with an increase in velocity. Consequently, PN predictions become
more and more
inaccurate the closer the binary gets to merger. Different analytical
modifications are known that try to enhance the convergence of the PN
series, even close to merger, and one of the most successful methods
is the \emph{effective-one-body} (EOB) approach
\cite{Buonanno:1998gg,Buonanno:2000ef,Damour:1997ub,Damour:2000we}.

Without further information, however, all these analytical schemes
break down at some point prior to the merger of both BHs, and
a
second approach has to be used to model the dynamics from the
late inspiral through the merger: \emph{numerical relativity} (NR).
In NR, the full Einstein equations are usually solved discretely on a
finite grid that is adapted to the movement of the two bodies,
and the resolution in space and time is chosen fine enough to obtain a
converging result.
 The GW content is extracted at finite radii and then extrapolated to
infinity, or it is directly extracted at null infinity via
Cauchy-characteristic
extraction~\cite{Reisswig:2009rx,Babiuc:2010ze}. For current
overviews of the field see for example
\cite{Hannam:2009rd,Hinder:2010vn,Centrella:2010zf,McWilliams:2010iq,
Sperhake:2011xk}.

\begin{figure}
 \centering
 \includegraphics[width=0.9\textwidth]{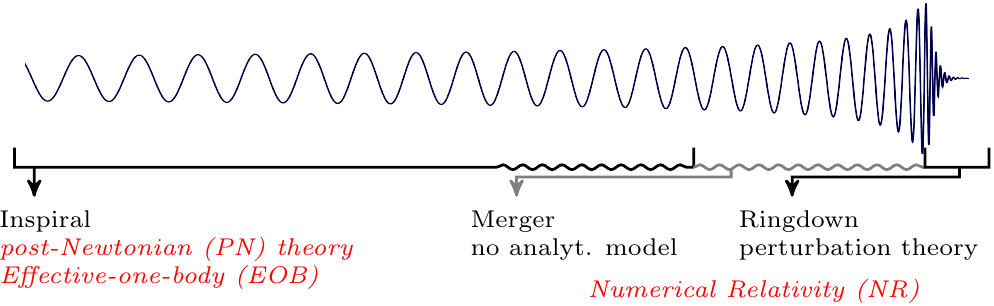}
 \caption{The dominant spherical harmonic mode of the gravitational
wave signal of two coalescing (nonspinning) BHs as a function of time.
The different approximation schemes and
their range of validity are indicated. Wavy lines illustrate the
regime close to merger where analytical methods have to be bridged by
NR.}
 \label{fig:longwave}
\end{figure}

Both numerical and analytical approaches have their limitations. The
PN-based formulations are, by construction, not valid throughout the
entire coalescence process; NR relies on computationally very
expensive simulations that become increasingly challenging (and
time-consuming) with larger initial separations, higher spin
magnitudes
of the BHs and higher mass-ratios $q = m_1/m_2$ ($m_i$ are the
masses of the individual BHs and we use the convention $m_1
\geq m_2$). Thus, to build models of the complete inspiral,
merger and ringdown signal, one has to combine information from both
analytical and numerical approximations. See Fig.~\ref{fig:longwave}
for an illustration of the dominant harmonic mode of a nonspinning
binary. 

These `complete' waveforms are indispensable to perfect current
search strategies. They constitute our best and most complete
approximation of the real signals that we are trying to detect, which
makes them ideal target waveforms in a simulated search to test
existing analysis algorithms. The Numerical INJection Analysis
(NINJA) project \cite{Aylott:2009ya,Ajith:2012tt} is dedicated to
that question. The other important application of complete waveforms
is to derive an analytical model from them which leads to an improved
template bank in the search. The improvement manifests itself, e.g., 
in a wider detection range and a more accurate extraction of the
physical information encoded in the signals. Ongoing 
searches with such templates in LIGO data are summarized for
instance in \cite{Abadie:2011kd}.

This paper briefly describes the efforts
to build complete waveform models by combining analytical approximants
and NR into individual signals and eventually entire
waveform families. Our focus then turns to the question of how
reliable
and accurate such final models are. After all, one expects (and finds)
a smooth connection between the two parts of a supposedly common GW
signal, but the use in actual analysis algorithms of GW
interferometers requires a much deeper error analysis with a
quantitative understanding of the uncertainty introduced in the
modelling process.

\section{Concepts for constructing full waveform models}
\label{sec:concepts}

\subsection{EOBNR}

The EOB formalism has been refined several times to incorporate
additional information from NR. Depending on the number of 
available NR waveforms as well as the modifications introduced to the
EOB description, various versions of such \emph{EOBNR} models have
been developed~\cite{Buonanno:2007pf,Buonanno:2009qa,Damour:2007vq,
Damour:2008te,Damour:2009kr,Pan:2009wj,Yunes:2009ef,Pan:2011gk}. 
It is beyond the scope of this paper to repeat the technical details
of the EOB formalism and its extensions. For the sake of comparison
to other approaches, however, we shall summarize the general strategy
towards complete inspiral-merger-ringdown EOBNR models below.

The main additions that allow for the description of the entire GW
signal are (a) a generalization of the EOB formalism which introduces
free parameters to be calibrated by NR simulations and (b) attaching
a series of damped sinusoidal oscillations (quasinormal modes)
representing the final stage of the BH ringdown (see, e.g.,
\cite{Berti:2009kk}). The proposed variants of EOBNR
mainly
differ in the way the original EOB description is modified
and which free parameters are introduced. The most recent versions by
Damour and Nagar \cite{Damour:2009kr} and Pan \emph{et
al.}~\cite{Pan:2011gk} extend the standard EOB form through the
following steps:
\begin{itemize}
 \item Two unknown parameters representing the 4PN- and 5PN-order
contributions are added to the radial potential [commonly referred to
as $A(u)$] that enters the Hamiltonian. As for many quantities in the
EOB framework, using \emph{Pad\'e} resummation
\cite{Damour:1997ub} proves to be
superior to the Taylor-expanded form (which is,
however, not always
true, see the discussion about a generalization to spinning BHs
\cite{Pan:2009wj} and also \cite{Mroue:2008fu}). 
 \item The radiation-reaction force and the waveform modes are
written in a resummed, factorized manner \cite{Damour:2008gu}.
Additional coefficients are introduced in the waveform, accounting
for further, undetermined PN contributions and next-to-quasi-circular
corrections.
 \item A sum of quasi-normal modes is attached to the inspiral-plunge
EOBNR waveform over a certain time interval around the peak of the
waveform mode. 
\end{itemize}

The impact of NR on the above strategy is manifold.
Some parameters (like the EOB-dynamical parameters introduced into the
radial potential) are directly determined through minimizing the
phase difference between the analytical and numerical GW. Other
parameters are derived from independent (i.e., not EOB-related) fits
of the numerical data, such as predictions of the final spin of the
remnant BH or the maximum of the modulus of the GW. 
Note, however, that for a direct comparison (and thereby calibration),
analytical and numerical waveforms have to be \emph{aligned},
i.e., a relative shift in time and phase has to be fixed by some
minimization procedure. We shall find the same need in all
construction algorithms for complete GW signals.

In short, the characteristics of EOBNR constructions are that a
well-adapted
analytical description is extended and \emph{informed} by NR data, so
that finally a time-domain description based on a set of differential
equations provides the entire inspiral to plunge signal that is
completed by attaching the ringdown waveform.

\subsection{Phenomenological models}

Although there is a common strategy in all modelling procedures
described here, let
us highlight a few distinct features of phenomenological waveform
families as introduced by Ajith \emph{et
al.}~\cite{Ajith:2007qp,Ajith:2007kx,Ajith:2009bn} and
Santamar\'ia \emph{et al.}~\cite{Santamaria:2010yb}.
These families are built by first constructing a finite set of
complete \emph{hybrid} waveforms
\cite{Ajith:2007kx,Boyle:2008ge,Boyle:2009dg,Santamaria:2010yb,
Hannam:2010ky, MacDonald:2011ne,Boyle:2011dy} that are direct
combinations
of the available NR data with the appropriate waveforms obtained with
some PN approximant (usually based
on Taylor-expanded quantities). 
The construction of these hybrid waveforms may differ, as they can be
based in the time or frequency domain, they can overlap both waveform
parts at a single point or over an interval and they can impose
various
requirements on the smoothness of the transition. However, all
hybridization procedures are based on finding the, in some sense,
optimal alignment between two parts of the same waveform by
exploiting the free relative time and phase shift. 

Different from the EOBNR approach, this combination of analytical and
numerical waveforms does not immediately lead to a model allowing
arbitrary physical parameters. 
Hybrid waveforms merely constitute the
set of discrete \emph{target signals} that are represented in a next
step as
accurately as possible by a simple and convenient multi-parameter fit.
This fit is separated from the analytical approach used
to describe the inspiral of the hybrid waveform. For instance, the
latest model of Ajith \emph{et al.}~\cite{Ajith:2009bn} employs a
time-domain PN approximant commonly denoted by `TaylorT1' in the
hybrid construction, but the final multi-parameter model is instead
inspired by the form of a Fourier-domain PN approximant (see for
instance \cite{Buonanno:2009zt} for an overview of the different PN
approximants). 

The final fit that turns a set of hybrid waveforms distributed in the
parameter space into an analytical model is a delicate procedure.
Introducing an arbitrary (yet as small as possible) number of
parameters to fit a relatively small number of hybrids is not
difficult. These auxiliary parameters, however, have to be a smooth
function of the \emph{physical parameters} (notably symmetric mass
ratio and spins) themselves in order to allow for an interpolation
of the parameter space. Only if the latter can be achieved, again
with guidance from PN descriptions and the knowledge of quasinormal
ringdown modes, the model becomes potentially useful for data-analysis
purposes without increasing the rate of false alarms in a
search process.

In the end, the phenomenological descriptions mentioned
above \cite{Ajith:2007qp,Ajith:2007kx,Ajith:2009bn,Santamaria:2010yb}
are provided in terms of closed-form equations representing the GW
signal in Fourier space. It should be noted that, although the
procedure of combining PN and NR data in a first step and analytically
modelling it in a second step is conceptually useful to analyse
different error sources (see Sec.~\ref{sec:errors}), it is not
entirely different
from the EOBNR approach. If the inspiral model used in the
hybrid would be EOB and an extended EOB description is chosen
as the `phenomenological model', then we would recover the EOBNR
construction.  Likewise, if the EOBNR construction would calibrate
its model against a complete hybrid signal instead of pure NR data, it
would be conceptually no different from phenomenological
constructions (which does not imply that one construction cannot be
superior to the other). The important question ultimately is how
\emph{flexible} and \emph{accurate} each individual strategy (with all
its detailed
distinctions) can predict the unknown real GW signal. We shall touch
this question in Sec.~\ref{sec:errors}.

For completeness, let us mention another phenomenological
family that was constructed by Sturani \emph{et al.}
\cite{Sturani:2010yv,Sturani:2010ju} as a first step to model
waveforms of precessing binaries. In this approach, a Taylor-expanded
time-domain approximant (`TaylorT4') is extended and finally fitted
to NR data. Just like EOBNR (although less sophisticated), the
resulting model is given in form of time-domain differential
equations with quasinormal ringdown modes attached.

\section{Physical range of waveform models}
 
Understanding the concepts underlying the construction of complete
waveform models is mainly interesting when we want to 
to compare various approaches, deduce why they lead to
slightly different waveforms and, most importantly, assess the
quality of individual families. In this section, however, we will
first
summarize the facts that are interesting for the actual usage of the
waveforms in data-analysis applications. In particular, before
applying the model to a set of physical parameters, one should have a
clear perception of \emph{where in the parameter space} these models
have
been constructed. Although this range does not necessarily coincide
with the range of parameters the model can be used with, it
nevertheless is a good indication where it can be trusted most.

The waveform models that have been introduced in
Section~\ref{sec:concepts} are tailored to model binary BHs
with
comparable masses inspiralling on quasi-circular orbits. There are
successful efforts to exploit the synergy of analytical methods and
NR also for other scenarios, like the extreme mass-ratio
regime \cite{Yunes:2009ef,Yunes:2010zj} or binary neutron star
coalescences \cite{Damour:2007vq,Baiotti:2011am}. In this paper, we
focus on binary BHs in the comparable-mass regime only,
as they are the most promising sources for the upcoming generation of
ground-based GW detectors whose detection and interpretation may
require information both from PN and NR. 

\fulltable{A selection of recent complete waveform models for
BH binaries with comparable masses on quasi-spherical
orbits. We summarize the reference where the model was
described, the approximate inspiral waveforms and NR codes that were
employed, the parameter range in which each model was calibrated ($q$
is the mass ratio) and the number of parameters and NR simulations
used to build the model. \label{tab:models}}
   \br
Alias & Ref. & Inspiral & NR code & Calibration range &
Calibrated parameters \\ 
\mr
EOBNR & \cite{Damour:2009kr} & EOB & \texttt{SpEC}, \texttt{BAM} & $q
\leq 4$ \& $q \to 0$ & 2 dynamical from $q=1$ \\ &&&&
no spins & + fits
from $q \in \{ 1,2,4 \}$ \\
EOBNR & \cite{Pan:2011gk} & EOB & \texttt{SpEC} & $q
\leq 6$ \& $q \to 0$ & 2 dyn. + 4 waveform par./mode \\ &&&&
no spins & 5 leading modes from 5 NR runs\\
Phenom\emph{B} & \cite{Ajith:2009bn} & T1 & \texttt{BAM} & $q
\leq 4$ \& $q \to 0$ & 6 phase, 4 amplitude \\ &&&&
aligned spins & from 24 NR simulations \\
Phenom\emph{C} & \cite{Santamaria:2010yb} & F2 & \texttt{BAM} & $q
\leq 4$  & 6 phase, 3 amplitude \\ &&&&
aligned spins & from 24 NR simulations \\
PhenSpin & \cite{Sturani:2010ju} & T4 & \texttt{MayaKranc} & $q
= 1, \chi_i = 0.6$  & 2 phase param. \\
&&&& precession & from 24 NR sim.+4 Phenom\emph{B}
\\ \br
\endfulltable

In Table~\ref{tab:models} we provide an overview of selected, recent
models for this regime.
Apart from an alias (partially
adopted from the LIGO-Virgo collaboration \cite{LSC,Virgo}) we
indicate the inspiral model which is either based on the EOB
approach or derived from Taylor-expanded PN quantities. In the latter
case, various
different PN approximants are known depending on the details of the
re-expansion and integration. The time-domain approximants are
commonly referred to as
\emph{TaylorTn} (where $n$ ranges from 1 to 4, and a fifth version
has recently been suggested \cite{Ajith:2011ec}); a 
frequency-domain representation obtained via the stationary-phase
approximation is denoted by \emph{TaylorF2}. For details, see
\cite{Boyle:2007ft,Buonanno:2009zt,Brown:2007jx} and references
therein. 

The NR codes that contributed to the construction of the given models
are the Spectral Einstein Code
(\texttt{SpEC}~\cite{Scheel:2006gg,Spec}),
\texttt{BAM}~\cite{Bruegmann:2006at} and
\texttt{MayaKranc}~\cite{Vaishnav:2007nm}, where the few \texttt{SpEC}
waveforms are notably long and accurate, the \texttt{BAM} simulations
provide the largest diversity in parameter space with moderately long
waveforms, and \texttt{MayaKranc} waveforms are the only precessing
simulations used to calibrate analytical models to date.

Other distinctive features of the models listed in
Table~\ref{tab:models} are for example as follows:
\begin{itemize}
 \item Phenom\emph{B}/\emph{C} are closed-form frequency-domain
representations of the GW; EOBNR and PhenSpin provide the signal in
terms of time-domain differential equations.
\item EOBNR models can readily be extended beyond the dominant
spherical harmonic of the GW, whereas the phenomenological models and
PhenSpin solely provide the signal in terms of the $\ell =2, m=\pm 2$
(spin-weighted) spherical harmonic modes.
\item The PhenSpin model is a first attempt to model generic
precessing
spin configurations, but it is so far only calibrated to equal-mass
systems and dimensionless spin magnitudes of 0.6.
All other models
in Table~\ref{tab:models} are only applicable to nonspinning systems
or systems where
the spin of each BH is aligned (or antialigned) with the total
orbital angular momentum. 
\end{itemize}
In the aligned-spin case, both 
phenomenological waveform families reduce the two spin parameters to
one ``total'' spin
\begin{equation}
 \chi = \frac{m_1 \chi_1 + m_2 \chi_2}{m_1 + m_2},
\end{equation}
where $m_i$ are the individual masses and the
dimensionless spin magnitudes are $\chi_i = \pm \vert \bi S _i 
\vert /m_i^2$ (the sign distinguishing aligned and antialigned
configurations).
 As recently shown
by Ajith~\cite{Ajith:2011ec}, this
degeneracy in the spin parameters can be further optimized, and it
will be an important goal for future models to describe
as many physical effects as possible with the smallest possible
number of parameters. In the nonspinning case, all waveforms presented
here are parametrized in terms of the \emph{physical} parameters
total mass and symmetric mass ratio (plus initial time and phase) but
it may be useful both from the modelling and the search point of
view to refrain from this parametrization strategy once all additional
spin
dynamics are included. Note that there is also an EOBNR model
proposed that includes aligned-spin configurations~\cite{Pan:2009wj},
but this first exploratory study only employed two equal-mass
simulations (performed with \texttt{SpEC}) with equal spins 
$\chi_1 = \chi_2 \approx \pm 0.44$. 

Apart from the listed facts, there are many more procedures involved 
in checking the validity of proposed models. Most importantly, it
has been shown to some extent that the models mentioned here agree to
reasonable accuracy with the waveforms they were derived from, but
also with waveforms that were \emph{not} in the construction set.
Thus, with an increasing number of available numerical simulations,
all these models can not only be extended and refined, they can also
be cross-checked extensively until, ideally, one can confidently
interpolate over  the entire parameter space independent of the
set of waveforms actually used to calibrate the model.

\section{Uncertainties in the modelling process} \label{sec:errors}

Having briefly sketched some successful approaches to
combine analytical and numerical methods in the GW-modelling process,
let us recapitulate the choices that had to be made along the
way. 
\begin{itemize}
 \item Which PN/EOB formulation should be employed? 
\item What \emph{physical parameters} in PN and NR
are consistent with the other framework?
 \item Which NR resolution, extraction formalism etc. is sufficient?
 \item How long do the NR waveforms have to be?
 \item What is the appropriate way to match analytical and numerical
data?
 \item How do the fitting parameters depend on physical
quantities?
\end{itemize}
When constructing a complete waveform model, each of these questions
has to be answered and different choices lead to the different
results presented above. The important conclusion we shall draw from
this is that none of the suggested models is based on an unambiguous
construction, and the spread of possible results that different
reasonable choices yield is a measure of the uncertainty within the
modelling process. We shall first outline the basic concepts of
evaluating these uncertainties in a way meaningful for GW searches
and then summarize some results that have been obtained in the recent
past.

\subsection{Accuracy requirements for detection and parameter
estimation}

Let us recall the basic strategy of a matched-filter search (see,
e.g.,
\cite{Finn:1992wt,Cutler:1994ys,lrr-2005-3}) that
employs a theoretically predicted template bank of waveforms
$h_2$ which in turn are characterized by the parameters $\theta$.
Assume that a real GW signal $h_1$ is contained in the data stream.
The search mainly relies on finding the maximum agreement
between $h_1$ and $h_2 (\theta)$, for any $\theta$, as
quantified by the inner product
\begin{equation}
 \big \langle h_1 , h_2 \big \rangle =  
4 \, {\rm Re} \int_{f_1}^{f_2} \frac{\tilde h_1(f) \, \tilde
h_2^\ast(f)}{S_n(f)} \, df. \label{eq:innerprod}
\end{equation}
Here, $\tilde h_i$ are the Fourier transforms of $h_i$, $^\ast$
denotes the complex conjugation, and $S_n$ is the noise spectral
density of the assumed GW detector (we assume stationary Gaussian
noise with zero mean).

The \emph{detection efficiency} can be expressed in
terms of the minimal mismatch
\begin{equation}
 \mathcal M = 1 - \max_{\theta} \frac{\big \langle  h_1,
h_2(\theta) \big \rangle}{\| h_1 \| \|h_2 (\theta) \|}, \qquad
(\| h_i \| = \sqrt{%
\big \langle h_i, h_i \big \rangle }), \label{eq:MM}
\end{equation}
which quantifies the loss in \emph{signal-to-noise ratio} (SNR) due to
an
inexact model, or equivalently yields the fraction of missed signals
in a matched-filter search. If, for example, up to $x=10\%$ of the
detectable signals may be missed due to a mismatch of real and
modelled waveform, we can allow this mismatch to be at most
$\mathcal M = 1  - \sqrt[3]{1-x} \approx  3.5\%$.

Typically, one does not have access to an ideal target waveform and an
approximate search family, so one commonly uses the mismatch between
two supposedly equivalent, approximate waveforms to quantify
the error of the modelling process itself. The other simplification
that often comes with the restriction to discrete points in the
parameter space is that instead of maximizing (\ref{eq:MM}) with
respect to all parameters, one only exploits a free relative time and
phase shift between the waveforms and varies along $\theta =
(t_0, \phi_0)$. This can only be an upper bound on $\mathcal M$, which
is of course sufficient if the values found are small enough.

The uncertainty of estimating parameters in a search has recently
\cite{Lindblom:2008cm,Damour:2010zb,Boyle:2011dy,MacDonald:2011ne}
been based  on the requirement
that the error of the waveform model is \emph{indistinguishable} by
the detector. With $h_1$ and $h_2$ denoting supposedly equivalent
waveforms, this requirement reads
\begin{equation}
\| \delta h \| = \| h_1 - h_2 \| < \epsilon, \label{eq:indist}
\end{equation}
with $\epsilon$ being the fraction of the
noise level we allow for the model uncertainty ($\epsilon \lesssim
1$). The
parameters of $h_1$ and $h_2$ are
deliberately kept the same in (\ref{eq:indist}), except for a time and
phase shift that is used to minimize the
distance. The other parameters (masses, spins) should be determined in
the search with no bias introduced by the model; thus, we do not
optimize over them in (\ref{eq:indist}). 
Note that (\ref{eq:indist}) can also be expressed in terms of the
mismatch (also minimized over a time and phase shift only) by assuming
the equal norms $\| h_1
\| = \| h_2 \| = \|h \|$ \cite{Flanagan:1997kp,McWilliams:2010eq},
\begin{equation}
\mathcal M = \frac12 \, \frac{\| \delta h \|^2}{\|h\|^2} \qquad
\Rightarrow \quad (\ref{eq:indist}) ~\Leftrightarrow~
 \mathcal M < \frac{\epsilon^2}{2 \| h \|^2}. \label{eq:misdistance}
\end{equation}

Fulfilling (\ref{eq:indist}) or (\ref{eq:misdistance}) is certainly
the ultimate goal
for an accurate waveform model, and we can easily understand that if
the waveform uncertainty $\delta h$ is not even detectable in the
presence of instrument noise, it has no effect on the
measurement. The converse, however, cannot be interpreted in this
straightforward manner. If
$\| \delta h \| > 1$, we would expect that the model uncertainty has
\emph{some} effect on the parameter estimation, but \emph{which}
effect it actually has must be quantified through the
\emph{systematic}
(i.e., model-induced) parameter bias which is defined
as the
difference between the parameter values of the best fitting template
and the true parameters \cite{Damour:1997ub}. These errors should then
be compared to statistical (i.e., noise-induced) errors, and there are
explicit expressions available for both types of biases in the high
SNR
regime \cite{Flanagan:1997kp,Cutler:2007mi}. Here we just remind the
reader that (\ref{eq:indist}) was derived as a \emph{sufficient}
criterion to ensure that the systematic errors
do not exceed the statistical parameter variance
\cite{Flanagan:1997kp}. It is, however, not a necessary criterion, and
we shall illustrate this explicitly in Sec.~\ref{sec:PNerr}.

Next, we summarize some important results that have been obtained
recently, all casting the question of accuracy of waveforms in
the form just outlined. 
All the
results quoted below use an Advanced LIGO noise curve
\cite{Ajith:2007kx,advLIGO} with
appropriate integration limits.

\subsection{Errors in the NR regime}

Quantifying errors is an important and very
natural process for numerical integrations and the uncertainty of NR
waveforms is usually given in terms of phase and amplitude
error. Although these can be related to the quantities we have
introduced above (see
\cite{Lindblom:2008cm,Lindblom:2009ux,Lindblom:2010mh}) we will focus
on publications here that directly analyse NR errors in terms of
waveform mismatches (all mismatches quoted below are optimized over
time and
phase shifts of the waveforms).
 
The `Samurai' project \cite{Hannam:2009hh} was a joint effort
proving the consistency of NR waveforms by comparing numerical
simulations of equal-mass, nonspinning binaries from five different NR
codes. No completion of the waveforms with PN or EOB inspiral
signals was considered as the focus laid primarily on the NR data and
their errors. Therefore, the mismatches reported in
\cite{Hannam:2009hh} are restricted to high
frequencies; thus, high masses of the system (total mass $> 180
M_\odot$), and values of $\mathcal M < 0.1\%$ are found. Similarly,
Santamar\'ia \emph{et al.}~\cite{Santamaria:2010yb} compare hybrid
waveforms of nonspinning binaries with mass ratios 1 and 2 where the
PN part was fixed, respectively, and NR data was produced either from
the \texttt{BAM} or \texttt{Llama}~\cite{Pollney:2009yz} code. The
maximal mismatch they find satisfies $\mathcal M < 0.2\%$.

The effect of different resolutions used to calculate the NR part of a
complete waveform was also analysed in \cite{Santamaria:2010yb} and
by MacDonald \emph{et al.}~\cite{MacDonald:2011ne} who report that
even the low-resolution run causes a difference to their best
simulation (nonspinning, equal mass, \texttt{SpEC} code) with 
$\mathcal M$ not greater than $0.1\%$.

Finally, various comparisons of analytical waveform models with NR
data can be found in the literature, e.g., the recent EOBNR model by
Pan \emph{et al.}~\cite{Pan:2011gk} exhibits a mismatch
$\mathcal M \sim 0.5\%$ with an
NR waveform (multiple harmonics) of a binary with mass ratio 6.
Smaller mass ratios and fewer harmonics lead to smaller
mismatches. Note that all these mismatches can be considered as small,
at least in terms of detection, as only $\approx 1.5\% ~(0.6\%)$ of
signals would be lost due to a mismatch of $\mathcal M = 0.5\%~
(0.2\%)$ (not even including additional optimizations over the
physical
parameters of the template bank).

\subsection{Hybridization errors}

There are different strategies to combine two parts of supposedly the
same waveform into one signal. As one is forced to
overlap analytical and numerical results in the late merger regime
(as early as the NR simulation permits) one cannot expect that they
agree perfectly, and there is some ambiguity about the way an
`optimal alignment' of both waveform parts is defined. Most common
both in EOBNR an phenomenological constructions is to decompose
$h$ into phase $\phi$ and amplitude $A$ by $h  = A \, e^{i \phi}$.
The alignment of the PN/EOB and NR parts is then carried out by
minimizing
a phase difference, either over an entire interval or at discrete
points, utilizing the frequency $\omega = d \phi/dt$ as well.

A detailed analysis of various aspects involved in such procedures
can be found in \cite{Santamaria:2010yb,MacDonald:2011ne}.
The authors of \cite{Santamaria:2010yb} calculate in their
example of a frequency-domain matching that the relative time shift
between both waveform parts can be determined, in the best case, up to
an uncertainty of $\delta t_0/M \approx 0.15$.
Ref.~\cite{MacDonald:2011ne} complements this statement by estimating
that $\delta t_0/M \lesssim 1$ is required for an accurate matching
with $\mathcal M < 0.02\%$. In addition, the recommended matching
interval is formulated in terms of the frequency evolution $\omega_1
\to \omega_2$ within this interval, and \cite{MacDonald:2011ne}
suggests $(\omega_2 - \omega_1) / \omega_m \gtrsim 0.1$
(where $\omega_m$ is
the transition frequency from PN to NR).

Hannam \emph{et al.}~\cite{Hannam:2010ky} compared different
hybridization schemes, including time and frequency-domain variants,
and they show for an equal-mass, nonspinning binary that the
resulting different hybrids have a mismatch of at most $\mathcal M =
0.03\%$. Again, we can summarize these results by stating that 
in the cases considered (mostly nonspinning, some aligned-spin
configurations, mass-ratio close to unity) the hybridization involves
some careful fine-tuning but the uncertainties introduced are
acceptable for data-analysis purposes.

\subsection{Uncertainty of the inspiral waveform -- NR length
requirements} \label{sec:PNerr}

Recently, several publications quantified the effect of different
analytical waveform models that are completed with common
late-inspiral, merger and ringdown
data~\cite{Santamaria:2010yb,Hannam:2010ky,Damour:2010zb,
MacDonald:2011ne,Boyle:2011dy,Ohme:2011zm}. The general approach in
each of these articles is similar: different PN/EOB approximants are
stitched to some given high-frequency data (Boyle~\cite{Boyle:2011dy}
and Ohme \emph{et al.}~\cite{Ohme:2011zm} point out that 
only very limited information from NR is actually needed) and the
slightly different signals are analysed in terms of their
distance $\| \delta h \|$ or mismatch $\mathcal M$, at first
optimized over time and phase shifts only.  

The results are sobering. Even approximants with
nonspinning/spinning terms up to 3.5PN/2.5PN order (amplitude at
3PN/2PN order) induce hybrid-waveform disagreements that can
reach mismatches of the order of a few to more than ten percent. That
is at least an order of magnitude more than what has been found for
NR and hybridization errors! 
Of course, there are again many details entering these
calculations, the most crucial choices being
(a) the two analytical models compared to each other,
(b) the total mass of the system,
(c) the other physical parameters
of the system (mass ratio, spins), and
(d) where the inspiral waveforms
are connected to common NR data.

In a conservative approach, (a) and
(b) are dealt with by repeating the same analysis
for various different models and calculating the mismatches for a
range of
masses (the waveforms themselves scale trivially with the total
mass); the maximum of all of these numbers then represents the total
uncertainty due to an ambiguous inspiral. However, astrophysical
expectations of, e.g., the minimal constituent mass may considerably
restrict the plausible range of total masses, which leads to more
relaxed accuracy requirements
particularly for higher mass ratios \cite{Boyle:2011dy,Ohme:2011zm}.

One way to reduce the total modelling error is to shorten the
PN contribution to the waveform and employ accordingly longer NR
simulations. As stated before, NR data are extremely
expensive to compute and therefore, estimating the required
minimal simulation length is a major input to be provided
by
the waveform modelling community. The first efforts in this direction
repeated the error analysis based on PN uncertainties
simply for different matching points, thereby estimating how early
one would have to match PN and NR to meet a given accuracy
requirement. The results found in
\cite{MacDonald:2011ne,Boyle:2011dy} suggest that, for given mass
ratio and spin, numerical waveforms have to be \emph{much longer}
than currently practical simulations. In particular, demanding an
uncertainty indistinguishable by the detector
$[$Eq.~(\ref{eq:indist})$]$, even for a moderate
SNR ($\sim 10$), requires \emph{hundreds of
NR orbits}.

At first sight, these results are discouraging, but they 
are based on sufficient (but not necessary) criteria for individual
waveforms. 
If instead the accuracy analysis is carried out for waveform families
that allow for a continuous variation of all physical parameters, the
corresponding error estimates
are much smaller.
Hannam \emph{et al.}~\cite{Hannam:2010ky} first presented mismatches
that are not only optimized over phase
and time shifts but also with respect to the total mass. Ohme
\emph{et al.}~\cite{Ohme:2011zm} calculated mismatches
that are fully optimized (also with respect to mass ratio and spin).
 This now allows us to quantify the errors not only
in terms
of waveform differences but also as 
modelling-based uncertainties in the determination of the parameters.
In practice, \cite{Ohme:2011zm}
reports fully optimized mismatches of the order of $1 \%$ and less,
achieved with parameter uncertainties of $1\%$ for total mass and
symmetric mass ratio and a total bias of $0.1$ for the spin parameter
$\chi$. It was concluded that approximately ten orbits before merger
of
numerical data should be good enough for the modelling of many
astrophysical systems.

The reason behind such inconsistent conclusions about the required NR
waveform length is that the authors take different approaches to
define an accuracy goal for hybrid waveforms. From a purely
theoretical point of view, every single waveform should be as
accurate as possible, and any effect on the measurement should be
excluded from the outset. This consequently leads to
(\ref{eq:indist}) and (\ref{eq:misdistance}) and the request for
hundreds of NR orbits. The other point of view is inspired by the
immediate goal to detect and interpret GWs out of existing data, and
although current waveform uncertainties significantly violate
(\ref{eq:indist}), it is concluded that one can still potentially use
waveforms incorporating $\approx$10 NR orbits for the science
intended. 

\begin{figure}
 \centering
 \includegraphics[width=0.43\textwidth]{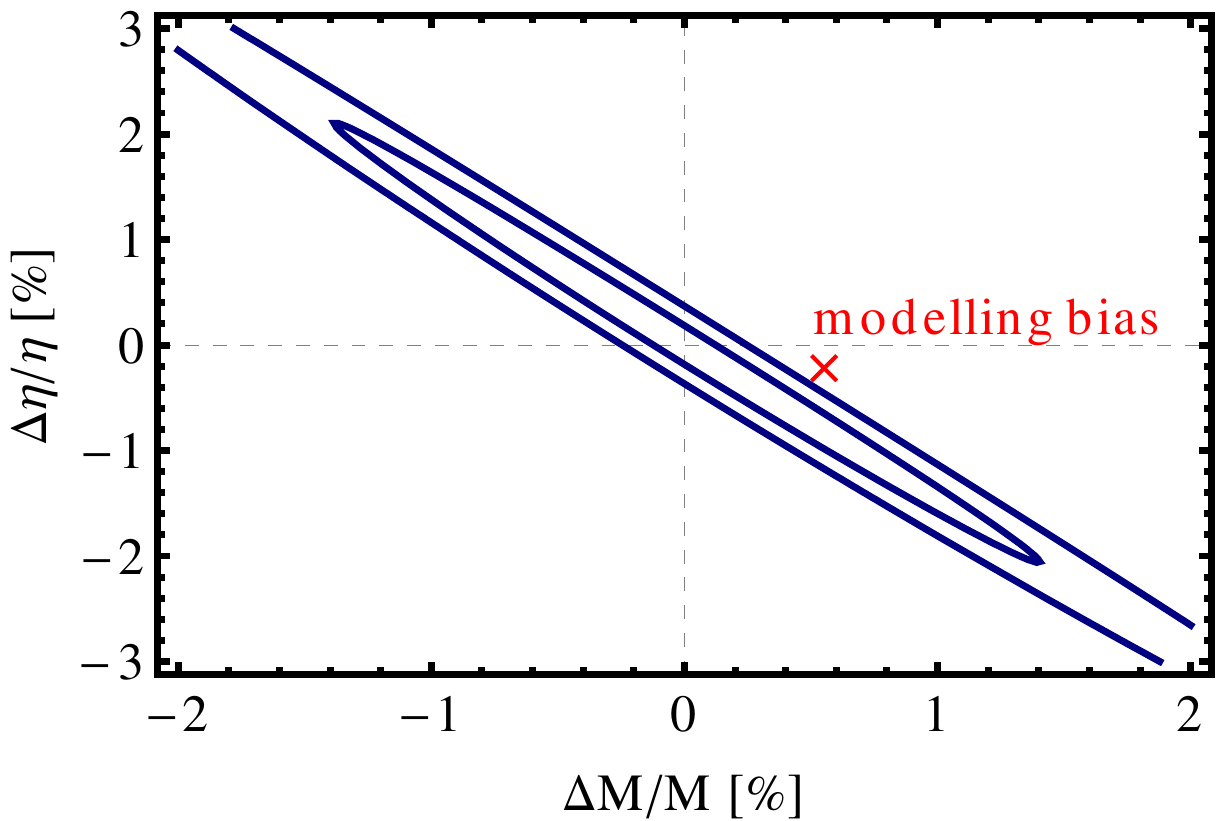}
\caption{The parameter uncertainties for a
binary system with $M=20M_\odot$, mass ratio 4 and $\chi = 0.5$. The
ellipses illustrate the statistical 1-$\sigma$ uncertainty for SNR 10
and 20 (inner ellipse). They are obtained through a calculation of the
Fisher information matrix (see \cite{Vallisneri:2007ev})
using the phenomenological model \cite{Santamaria:2010yb} with all its
parameters. The systematic modelling bias is adopted from
\cite{Ohme:2011zm}, and we use the same detector configuration here,
i.e., the analytical fit of the design sensitivity for Advanced LIGO
\cite{Ajith:2007kx} with a lower cutoff frequency at 20 Hz.}
\label{fig:Biases}
\end{figure}

This much more optimistic conclusion relies on fully optimized
mismatches, but also on the actual calculation of the systematic
parameter bias. As we have noted before, not satisfying
(\ref{eq:indist}) does not imply that the systematic bias exceeds the
statistical one, and an illustration of that is given in
Fig.~\ref{fig:Biases}. For the example of a binary with total mass $M
= 20M_\odot$, mass ratio 4
and equal aligned spins with $\chi = 0.5$, we see that the modelling
bias calculated in \cite{Ohme:2011zm} is outside of the 1-$\sigma$
ellipse (by which we mean the area
that satisfies (\ref{eq:misdistance}) with $\epsilon = 1$) for all
reasonable SNRs. The
uncertainty of measuring the total mass $M$ and the symmetric
mass-ratio $\eta$ is nevertheless dominated by the statistical error
for moderate SNRs, as we can infer from the extension of the ellipses.
(If we would measure the chirp mass $M_c = M \eta^{3/5}$ instead of
the total mass, we would find the opposite relation, which highlights
again that there is some non-negligible effect on measuring
parameters, but it has to be interpreted carefully.)

In the end, both views are important to assess the current state of
the art in building GW models. Neither are 10 NR orbits good enough
for every application nor are currently feasible waveform models
entirely useless for detection or parameter estimation purposes.

\subsection{Interpolation error}

The final source of error we shall discuss here is the step from a
discrete set of combined PN+NR waveforms to an analytical model. This
involves a choice of the interpolation model that eventually
represents the entire (phenomenological) waveform as a function of the
physical parameters, or, less obvious in the EOBNR construction, the
concrete dependence of certain parameters has to be fixed by hand
(see, e.g., Fig.~5 in \cite{Pan:2011gk}). Investigations of how
different choices affect the entire model have mostly been restricted
to comparisons with NR data; mismatches of the entire signals in
different points of the parameter space, however, have not been
published to the extent other error sources have been analysed.

An indication of how relevant these `interpolation errors' are
is provided by the study of Damour, Nagar and Trias
\cite{Damour:2010zb} who compared an EOBNR model \cite{Damour:2009kr}
with phenomenological models
\cite{Ajith:2007kx,Ajith:2009bn,Santamaria:2010yb}, showing that even
the mismatches optimized over physical parameters (excluding the spin)
exceed 3\% in some regions of the parameter space. At first sight,
this might be surprising as the hybrids used to construct the models
should be accurate enough for detection purposes (satisfying the 3\%
mismatch criterion). The difference between the final model and
hybrids is also reported to be $\mathcal M \lesssim 2\%$ ($\lesssim
5\%$ for the PhenSpin model). 

It should be noted, however, that the triangle inequality reads
\begin{equation}
 \| h_{\rm model} - h_{\rm exact} \| \leq \| h_{\rm model} - h_{\rm
hybrid} \| + \| h_{\rm hybrid} - h_{\rm exact} \|,
\end{equation}
which yields through the relation (\ref{eq:misdistance}) and its
assumptions 
\begin{equation}
 \mathcal M (h_{\rm model}, h_{\rm exact} ) \leq \left( \sqrt{\mathcal
M (h_{\rm model} , h_{\rm hybrid})} + \sqrt{\mathcal M (h_{\rm
hybrid}, h_{\rm exact})} \right)^2. \label{eq:mm_triagle}
\end{equation}
Consequently, if the hybrids are accurate within, say, 2\%
mismatch and the model does not deviate by more than 2\% from the
set of hybrids, the resulting total uncertainty can nevertheless only
be bounded to 8\%, which is far above the acceptable mismatch. It is
clear from this rough estimation and the results from
\cite{Damour:2010zb} that the interpolation of the final model has to
be improved in the future, which can be done most easily by increasing
the number of (NR/hybrid) waveforms it is constructed from.

\section{Summary}

Because of the rapid advance in numerical relativity and analytical
waveform modelling, there are already a number of waveform models
proposed that describe the entire inspiral, merger and ringdown
signal of a BH binary. These models are already used to
analyse data from GW laser interferometers \cite{Abadie:2011kd} and
their impact will
successively grow the more physical effects are understood and
included in the model. One of the upcoming challenges is for instance
the improved construction of full waveform models with precessing
spins, and some fundamental questions concerning the choice of
coordinate system have already been
addressed~\cite{Schmidt:2010it,O'Shaughnessy:2011fx,Boyle:2011gg,
O'Shaughnessy:2012vm}.

Some attention of the modelling community has most recently
been on investigating different error sources and ambiguities in the
modelling process, and it was shown that different waveform
models do not yet agree accurately enough so that their uncertainty
can be neglected. Of course, current and future studies (such as
the NINJA project \cite{Aylott:2009ya,Ajith:2012tt} and the NRAR
collaboration) aim
at testing various search algorithms and waveform models, and the
`best model' will subsequently be refined by incorporating the
results obtained in such analyses. It should be pointed out, however,
that the variety of models we have today is very useful. In fact,
most of the quantitative error analyses rely on the diversity
different approaches generate, and if they eventually converge to an
(almost) unambiguous description of the waveform, this will put GW
astronomy on very solid ground.

The prospects for that are rather good. Techniques to
accurately simulate compact binaries and extract GWs are
advancing, and the more efficient NR codes become the more they can
provide large sets of waveforms spanning the parameter space. A few
very long, very accurate simulations will greatly aid the analysis of
fundamental questions related to the combination of analytical and
numerical data, but there are already various established and
well-tested strategies to match both waveform parts. In addition,
most recent studies show that hybrid waveforms employing moderately
long NR signals are already potentially useful for data-analysis
applications, so refining existing models or introducing new
complete waveform models on the basis of many of those NR waveforms
should be a realistic goal to be accomplished before the advent of
the advanced-detector era.

The common framework to quantify the uncertainty in various waveform
models has been introduced and used by several authors, which not
only allows for a meaningful comparison of different approaches, it
also sets the stage for the time when actually measured signals have
to be interpreted. The question of how confident one can claim to
identify the source of the signal will be of fundamental importance,
and any signal that exceeds the uncertainty limits of all
theoretically modelled waveforms will be equally exciting. 

\ack

It is a pleasure to thank the organizers at Cardiff University for a
very pleasant and enlightening conference NRDA2011/Amaldi 9. Many
ideas and insights presented here are the result of numerous
discussions with colleagues, with special thanks to Mark Hannam,
Sascha Husa, Badri Krishnan, Stas Babak and Parameswaran Ajith. This
work was supported by the IMPRS for Gravitational Wave Astronomy.

\footnotesize

\setlength{\bibsep}{0pt}

\bibliographystyle{iopart-num}
\bibliography{NRDA2011Proc}

\providecommand{\newblock}{}
\begin{thebibliography}{10}
\expandafter\ifx\csname url\endcsname\relax
  \def\url#1{{\tt #1}}\fi
\expandafter\ifx\csname urlprefix\endcsname\relax\def\urlprefix{URL }\fi
\providecommand{\eprint}[2][]{\url{#2}}

\bibitem{Abbott:2007kv}
Abbott B {\em et~al.\/} (LIGO Scientific Collaboration) 2009 {\em
  Rept.Prog.Phys.\/} {\bf 72} 076901 (\textit{Preprint} \eprint{0711.3041})

\bibitem{Sigg:2008zz}
Sigg D (LIGO Scientific Collaboration) 2008 {\em Class.Quant.Grav.\/} {\bf 25}
  114041

\bibitem{Smith:2009bx}
Smith J~R (LIGO Scientific Collaboration) 2009 {\em Class.Quant.Grav.\/} {\bf
  26} 114013 (\textit{Preprint} \eprint{0902.0381})

\bibitem{Acernese:2008zzf}
Acernese F, Alshourbagy M, Amico P, Antonucci F, Aoudia S {\em et~al.\/} 2008
  {\em Class.Quant.Grav.\/} {\bf 25} 184001

\bibitem{Accadia:2011zz}
Accadia T, Acernese F, Antonucci F, Astone P, Ballardin G {\em et~al.\/} 2011
  {\em Class.Quant.Grav.\/} {\bf 28} 114002

\bibitem{Grote:2008zz}
Grote H (LIGO Scientific Collaboration) 2008 {\em Class.Quant.Grav.\/} {\bf 25}
  114043

\bibitem{Luck:2010rs}
Luck H (LIGO Scientific Collaboration) 2010  (\textit{Preprint}
  \eprint{1004.0338})

\bibitem{Abadie:2010cf}
Abadie J {\em et~al.\/} (LIGO Scientific Collaboration, Virgo Collaboration)
  2010 {\em Class.Quant.Grav.\/} {\bf 27} 173001 (\textit{Preprint}
  \eprint{1003.2480})

\bibitem{lrr-2006-4}
Blanchet L 2006 {\em Living Reviews in Relativity\/} {\bf 9} 4
  \urlprefix\url{http://www.livingreviews.org/lrr-2006-4}

\bibitem{Buonanno:1998gg}
Buonanno A and Damour T 1999 {\em Phys. Rev.\/} {\bf D59} 084006
  (\textit{Preprint} \eprint{gr-qc/9811091})

\bibitem{Buonanno:2000ef}
Buonanno A and Damour T 2000 {\em Phys. Rev.\/} {\bf D62} 064015
  (\textit{Preprint} \eprint{gr-qc/0001013})

\bibitem{Damour:1997ub}
Damour T, Iyer B~R and Sathyaprakash B~S 1998 {\em Phys. Rev.\/} {\bf D57}
  885--907 (\textit{Preprint} \eprint{gr-qc/9708034})

\bibitem{Damour:2000we}
Damour T, Jaranowski P and Sch{\"a}fer G 2000 {\em Phys. Rev.\/} {\bf D62}
  084011 (\textit{Preprint} \eprint{gr-qc/0005034})

\bibitem{Reisswig:2009rx}
Reisswig C, Bishop N, Pollney D and Szilagyi B 2010 {\em Class.Quant.Grav.\/}
  {\bf 27} 075014 (\textit{Preprint} \eprint{0912.1285})

\bibitem{Babiuc:2010ze}
Babiuc M, Szilagyi B, Winicour J and Zlochower Y 2011 {\em Phys.Rev.\/} {\bf
  D84} 044057 (\textit{Preprint} \eprint{1011.4223})

\bibitem{Hannam:2009rd}
Hannam M 2009 {\em Class. Quant. Grav.\/} {\bf 26} 114001 (\textit{Preprint}
  \eprint{0901.2931})

\bibitem{Hinder:2010vn}
Hinder I 2010 {\em Class. Quant. Grav.\/} {\bf 27} 114004 (\textit{Preprint}
  \eprint{1001.5161})

\bibitem{Centrella:2010zf}
Centrella J~M, Baker J~G, Kelly B~J and van Meter J~R 2010 {\em
  Ann.Rev.Nucl.Part.Sci.\/} {\bf 60} 75--100 (\textit{Preprint}
  \eprint{1010.2165})

\bibitem{McWilliams:2010iq}
McWilliams S~T 2011 {\em Class.Quant.Grav.\/} {\bf 28} 134001
  (\textit{Preprint} \eprint{1012.2872})

\bibitem{Sperhake:2011xk}
Sperhake U, Berti E and Cardoso V 2011  (\textit{Preprint} \eprint{1107.2819})

\bibitem{Aylott:2009ya}
Aylott B {\em et~al.\/} 2009 {\em Class.Quant.Grav.\/} {\bf 26} 165008
  (\textit{Preprint} \eprint{0901.4399})

\bibitem{Ajith:2012tt}
Ajith P, Boyle M, Brown D~A, Brugmann B, Buchman L~T {\em et~al.\/} 2012
  (\textit{Preprint} \eprint{1201.5319})

\bibitem{Abadie:2011kd}
Abadie J {\em et~al.\/} (The LIGO Scientific Collaboration and the Virgo
  Collaboration, the Virgo Collaboration) 2011 {\em Phys.Rev.\/} {\bf D83}
  122005 (\textit{Preprint} \eprint{1102.3781})

\bibitem{Buonanno:2007pf}
Buonanno A {\em et~al.\/} 2007 {\em Phys. Rev.\/} {\bf D76} 104049
  (\textit{Preprint} \eprint{0706.3732})

\bibitem{Buonanno:2009qa}
Buonanno A {\em et~al.\/} 2009 {\em Phys. Rev.\/} {\bf D79} 124028
  (\textit{Preprint} \eprint{0902.0790})

\bibitem{Damour:2007vq}
Damour T, Nagar A, Dorband E~N, Pollney D and Rezzolla L 2008 {\em Phys.
  Rev.\/} {\bf D77} 084017 (\textit{Preprint} \eprint{0712.3003})

\bibitem{Damour:2008te}
Damour T, Nagar A, Hannam M, Husa S and Br{\"u}gmann B 2008 {\em Phys. Rev.\/}
  {\bf D78} 044039 (\textit{Preprint} \eprint{0803.3162})

\bibitem{Damour:2009kr}
Damour T and Nagar A 2009 {\em Phys. Rev.\/} {\bf D79} 081503
  (\textit{Preprint} \eprint{0902.0136})

\bibitem{Pan:2009wj}
Pan Y {\em et~al.\/} 2010 {\em Phys. Rev.\/} {\bf D81} 084041
  (\textit{Preprint} \eprint{0912.3466})

\bibitem{Yunes:2009ef}
Yunes N, Buonanno A, Hughes S~A, Coleman~Miller M and Pan Y 2010 {\em Phys.
  Rev. Lett.\/} {\bf 104} 091102 (\textit{Preprint} \eprint{0909.4263})

\bibitem{Pan:2011gk}
Pan Y {\em et~al.\/} 2011 {\em Phys.Rev.\/} {\bf D84} 124052 (\textit{Preprint}
  \eprint{1106.1021})

\bibitem{Berti:2009kk}
Berti E, Cardoso V and Starinets A~O 2009 {\em Class.Quant.Grav.\/} {\bf 26}
  163001 (\textit{Preprint} \eprint{0905.2975})

\bibitem{Mroue:2008fu}
Mrou{\'e} A~H, Kidder L~E and Teukolsky S~A 2008 {\em Phys.Rev.\/} {\bf D78}
  044004 (\textit{Preprint} \eprint{0805.2390})

\bibitem{Damour:2008gu}
Damour T, Iyer B~R and Nagar A 2009 {\em Phys.Rev.\/} {\bf D79} 064004
  (\textit{Preprint} \eprint{0811.2069})

\bibitem{Ajith:2007qp}
Ajith P {\em et~al.\/} 2007 {\em Class. Quant. Grav.\/} {\bf 24} S689--S700
  (\textit{Preprint} \eprint{0704.3764})

\bibitem{Ajith:2007kx}
Ajith P {\em et~al.\/} 2008 {\em Phys. Rev.\/} {\bf D77} 104017
  (\textit{Preprint} \eprint{0710.2335})

\bibitem{Ajith:2009bn}
Ajith P {\em et~al.\/} 2011 {\em Phys. Rev. Lett.\/} {\bf 106} 241101
  (\textit{Preprint} \eprint{0909.2867})

\bibitem{Santamaria:2010yb}
Santamar{\'i}a L {\em et~al.\/} 2010 {\em Phys. Rev.\/} {\bf D82} 064016
  (\textit{Preprint} \eprint{1005.3306})

\bibitem{Boyle:2008ge}
Boyle M {\em et~al.\/} 2008 {\em Phys. Rev.\/} {\bf D78} 104020
  (\textit{Preprint} \eprint{0804.4184})

\bibitem{Boyle:2009dg}
Boyle M, Brown D~A and Pekowsky L 2009 {\em Class. Quant. Grav.\/} {\bf 26}
  114006 (\textit{Preprint} \eprint{0901.1628})

\bibitem{Hannam:2010ky}
Hannam M, Husa S, Ohme F and Ajith P 2010 {\em Phys.Rev.\/} {\bf D82} 124052
  (\textit{Preprint} \eprint{1008.2961})

\bibitem{MacDonald:2011ne}
MacDonald I, Nissanke S and Pfeiffer H~P 2011 {\em Class.Quant.Grav.\/} {\bf
  28} 134002 (\textit{Preprint} \eprint{1102.5128})

\bibitem{Boyle:2011dy}
Boyle M 2011 {\em Phys.Rev.\/} {\bf D84} 064013 (\textit{Preprint}
  \eprint{1103.5088})

\bibitem{Buonanno:2009zt}
Buonanno A, Iyer B, Ochsner E, Pan Y and Sathyaprakash B~S 2009 {\em Phys.
  Rev.\/} {\bf D80} 084043 (\textit{Preprint} \eprint{0907.0700})

\bibitem{Sturani:2010yv}
Sturani R {\em et~al.\/} 2010 {\em J. Phys. Conf. Ser.\/} {\bf 243} 012007
  (\textit{Preprint} \eprint{1005.0551})

\bibitem{Sturani:2010ju}
Sturani R, Fischetti S, Cadonati L, Guidi G, Healy J {\em et~al.\/} 2010
  (\textit{Preprint} \eprint{1012.5172})

\bibitem{Yunes:2010zj}
Yunes N {\em et~al.\/} 2011 {\em Phys.Rev.\/} {\bf D83} 044044
  (\textit{Preprint} \eprint{1009.6013})

\bibitem{Baiotti:2011am}
Baiotti L, Damour T, Giacomazzo B, Nagar A and Rezzolla L 2011 {\em
  Phys.Rev.\/} {\bf D84} 024017 (\textit{Preprint} \eprint{1103.3874})

\bibitem{LSC}
{LIGO Scientific Collaboration} \url{http://www.ligo.org}

\bibitem{Virgo}
{Virgo Collaboration} \url{https://wwwcascina.virgo.infn.it}

\bibitem{Ajith:2011ec}
Ajith P 2011 {\em Phys.Rev.\/} {\bf D84} 084037 (\textit{Preprint}
  \eprint{1107.1267})

\bibitem{Boyle:2007ft}
Boyle M {\em et~al.\/} 2007 {\em Phys. Rev.\/} {\bf D76} 124038
  (\textit{Preprint} \eprint{0710.0158})

\bibitem{Brown:2007jx}
Ajith P {\em et~al.\/} 2007  (\textit{Preprint} \eprint{0709.0093})

\bibitem{Scheel:2006gg}
Scheel M~A {\em et~al.\/} 2006 {\em Phys.Rev.\/} {\bf D74} 104006
  (\textit{Preprint} \eprint{gr-qc/0607056})

\bibitem{Spec}
{The Spectral Einstein Code (SpEC)} \url{http://www.black-holes.org/SpEC.html}

\bibitem{Bruegmann:2006at}
Br{\"u}gmann B {\em et~al.\/} 2008 {\em Phys.Rev.\/} {\bf D77} 024027
  (\textit{Preprint} \eprint{gr-qc/0610128})

\bibitem{Vaishnav:2007nm}
Vaishnav B, Hinder I, Herrmann F and Shoemaker D 2007 {\em Phys. Rev.\/} {\bf
  D76} 084020 (\textit{Preprint} \eprint{0705.3829})

\bibitem{Finn:1992wt}
Finn L~S 1992 {\em Phys.Rev.\/} {\bf D46} 5236--5249 (\textit{Preprint}
  \eprint{gr-qc/9209010})

\bibitem{Cutler:1994ys}
Cutler C and Flanagan E~E 1994 {\em Phys.Rev.\/} {\bf D49} 2658--2697
  (\textit{Preprint} \eprint{gr-qc/9402014})

\bibitem{lrr-2005-3}
Jaranowski P and Kr{\'o}lak A 2005 {\em Living Reviews in Relativity\/} {\bf 8}
  3 \urlprefix\url{http://www.livingreviews.org/lrr-2005-3}

\bibitem{Lindblom:2008cm}
Lindblom L, Owen B~J and Brown D~A 2008 {\em Phys. Rev.\/} {\bf D78} 124020
  (\textit{Preprint} \eprint{0809.3844})

\bibitem{Damour:2010zb}
Damour T, Nagar A and Trias M 2011 {\em Phys.Rev.\/} {\bf D83} 024006
  (\textit{Preprint} \eprint{1009.5998})

\bibitem{Flanagan:1997kp}
Flanagan E~E and Hughes S~A 1998 {\em Phys.Rev.\/} {\bf D57} 4566--4587
  (\textit{Preprint} \eprint{gr-qc/9710129})

\bibitem{McWilliams:2010eq}
McWilliams S~T, Kelly B~J and Baker J~G 2010 {\em Phys.Rev.\/} {\bf D82} 024014
  (\textit{Preprint} \eprint{1004.0961})

\bibitem{Cutler:2007mi}
Cutler C and Vallisneri M 2007 {\em Phys.Rev.\/} {\bf D76} 104018
  (\textit{Preprint} \eprint{0707.2982})

\bibitem{advLIGO}
Shoemaker D 2009 {Advanced LIGO anticipated sensitivity curves}
  \url{https://dcc.ligo.org/cgi-bin/ DocDB/ShowDocument?docid=2974}

\bibitem{Lindblom:2009ux}
Lindblom L 2009 {\em Phys.Rev.\/} {\bf D80} 064019 (\textit{Preprint}
  \eprint{0907.0457})

\bibitem{Lindblom:2010mh}
Lindblom L, Baker J~G and Owen B~J 2010 {\em Phys.Rev.\/} {\bf D82} 084020
  (\textit{Preprint} \eprint{1008.1803})

\bibitem{Hannam:2009hh}
Hannam M {\em et~al.\/} 2009 {\em Phys. Rev.\/} {\bf D79} 084025
  (\textit{Preprint} \eprint{0901.2437})

\bibitem{Pollney:2009yz}
Pollney D {\em et~al.\/} 2011 {\em Phys.Rev.\/} {\bf D83} 044045
  (\textit{Preprint} \eprint{0910.3803})

\bibitem{Ohme:2011zm}
Ohme F, Hannam M and Husa S 2011 {\em Phys.Rev.\/} {\bf D84} 064029
  (\textit{Preprint} \eprint{1107.0996})

\bibitem{Vallisneri:2007ev}
Vallisneri M 2008 {\em Phys.Rev.\/} {\bf D77} 042001 (\textit{Preprint}
  \eprint{gr-qc/0703086})

\bibitem{Schmidt:2010it}
Schmidt P, Hannam M, Husa S and Ajith P 2011 {\em Phys.Rev.\/} {\bf D84} 024046
  (\textit{Preprint} \eprint{1012.2879})

\bibitem{O'Shaughnessy:2011fx}
O'Shaughnessy {\em et~al.\/} 2011 {\em Phys.Rev.\/} {\bf D84} 124002
  (\textit{Preprint} \eprint{1109.5224})

\bibitem{Boyle:2011gg}
Boyle M, Owen R and Pfeiffer H~P 2011 {\em Phys.Rev.\/} {\bf D84} 124011
  (\textit{Preprint} \eprint{1110.2965})

\bibitem{O'Shaughnessy:2012vm}
O'Shaughnessy R, Healy J, London L, Meeks Z and Shoemaker D 2012
  (\textit{Preprint} \eprint{1201.2113})

\end{thebibliography}

\end{document}